\documentclass[9pt,journal]{IEEEtran}

\usepackage[noadjust]{cite}

\usepackage{hyperref}
\newcommand{\figref}[2][{}]{\hyperref[#2]{\figurename~\ref{#2}#1}}

\ifCLASSINFOpdf
   \usepackage[pdftex]{graphicx}
   \graphicspath{{./pdf/}{./jpeg/}}
   \DeclareGraphicsExtensions{.pdf,.jpeg,.png}
\else
   \usepackage[dvips]{graphicx}
   \graphicspath{{../eps/}}
   \DeclareGraphicsExtensions{.eps}
\fi

\usepackage{amsmath}
\usepackage{amssymb}
\usepackage{cleveref}
\usepackage{booktabs}
\usepackage{tabu}
\usepackage{longtable}
\usepackage{bm}
\usepackage{multirow}

\usepackage{arydshln}

\usepackage{array}

\ifCLASSOPTIONcompsoc
 \usepackage[caption=false,font=normalsize,labelfont=sf,textfont=sf]{subfig}
\else
 \usepackage[caption=false,font=footnotesize]{subfig}
\fi

\usepackage{stfloats}

\ifCLASSOPTIONcaptionsoff
 \usepackage[nomarkers]{endfloat}
\let\MYoriglatexcaption\caption
\renewcommand{\caption}[2][\relax]{\MYoriglatexcaption[#2]{#2}}
\fi

\usepackage{listings}
\usepackage{courier}

\usepackage{url}

\begin{document}
\title{Mass-Matrix Differential-Algebraic Equation Formulation for Transient Stability Simulation}

\author{Hantao~Cui,~\IEEEmembership{Member,~IEEE,}
        Fangxing~(Fran)~Li,~\IEEEmembership{Fellow,~IEEE,}
        Joe~H.~Chow,~\IEEEmembership{Life Fellow,~IEEE}%
\thanks{H. Cui and F. Li are with the Department
of Electrical Engineering and Computer Science, The University of Tennessee, Knoxville,
TN, 37996 USA e-mail: fli6@utk.edu.}%
\thanks{J. H. Chow is with the Department of Electrical, 
Computer, and Systems Engineering, 
Rensselaer Polytechnic Institute, Troy, NY 12180 USA}
\thanks{This work was supported in part by the Engineering Research Center Program of the National Science Foundation and the Department of Energy under NSF Award Number EEC-1041877 and the CURENT Industry Partnership Program.}%
}%
\markboth{Preprint to be submitted to IEEE Power Engineering Society Letters}%
{Cui \MakeLowercase{\textit{et al.}}: Mass-Matrix Differential}

\maketitle

\begin{abstract}
This letter proposes a mass-matrix differential-algebraic equation (DAE) formulation for transient stability simulation. 
This formulation has two prominent advantages: compatible with a multitude of implicit DAE solvers and can be conveniently implemented based on the traditional formulation, 
for example, 
by separating the parameters in denominators into the diagonals of the mass matrix. 
It also allows reducing the dynamics using null time constants.
Benchmark studies are presented on the time and accuracy of 17 implicit solvers for the proposed formulation using the Kundur's two-area system and a 2,000 bus system.
\end{abstract}

\IEEEpeerreviewmaketitle

\section{Introduction}
\IEEEPARstart{P}{ower} systems transient stability programs traditionally employ the following explicit differential-algebraic equations (DAEs) for transient stability analysis:
  \begin{equation}
    \label{eq:dae-traditional}
    \begin{array}{lll}
      \dot{\bm{x}} &=& \bm{f}(\bm{x}, \bm{y}, \bm{u})  \\
      \bm{0} &=& \bm{g}(\bm{x}, \bm{y}, \bm{u})  \, , \\
    \end{array}
  \end{equation}
 where
 $\bm{f}$ 
 ($\bm{f}: \mathbb{R}^{m+n+o} \Rightarrow \mathbb{R}^n$)
 and $\bm{g}$
 ($\bm{g}: \mathbb{R}^{m+n+o} \Rightarrow \mathbb{R}^m$)
 are the differential and the algebraic equations, respectively, $\bm{x}$ are the differential states, $\bm{y}$ are the algebraic variables, and $\bm{u}$ are discontinuous states from discrete models.
 
 Recent works such as \cite{aristidou2013dynamic} propose an explicit formulation that can convert differential states into algebraic ones. This is referred to as the flexible formulation given by
\begin{equation}
    \label{eq:dae-flexible}
    \begin{array}{rll}
      \bm{\Gamma} \dot{\bm{x}} &=& \bm{f}(\bm{x}, \bm{y}, \bm{u})  \\
      \bm{0} &=& \bm{g}(\bm{x}, \bm{y}, \bm{u})  \, , \\
    \end{array}
\end{equation}
where $\bm{\Gamma}$ is a $n \times n$ time-invariant diagonal matrix with diagonal $\gamma_{i, i}$ satisfying
\begin{equation}
\left\{\begin{matrix}
\gamma_{i, i}=1, & \text{if } x_i \text{ remains as a differential state,} \\ 
\gamma_{i, i}=0, & \text{if } x_i \text{ is converted to an algebraic one.}
\end{matrix}\right.
\end{equation}

More recently, a semi-implicit formulation is proposed in 
\cite{milano2016semi} 
with the abilities to 1) increase the sparsity of Jacobians, 2) reduce computation efforts, and 3) enable state-to-algebraic switching, given by
\begin{equation}
    \label{eq:dae-semi-implicit}
    \begin{array}{lll}
      \bm{T}(\bm{x},\bm{y}) \dot{\bm{x}} &=& \bm{\tilde{f}}(\bm{x}, \bm{y}, \bm{u})  \\
      \bm{R}(\bm{x},\bm{y}) \dot{\bm{x}} &=& \bm{\tilde{g}}(\bm{x}, \bm{y}, \bm{u})  \, , \\
    \end{array}
\end{equation}
where $\bm{T}(\bm{x},\bm{y})$ and $\bm{R}(\bm{x},\bm{y})$ are $n\times n$ and $m \times n$ time-variant, highly sparse, non-diagonal and non-full rank matrices.
This formulation, however, can be challenging to implement due to the complexity and effort associated with implementing custom solvers and porting models from \eqref{eq:dae-traditional} to \eqref{eq:dae-semi-implicit}, which, although, can be carried out incrementally.

This letter proposes a mass-matrix formulation that can take advantage of numerous DAE solvers and is straightforward to implement.
It also allows for converting differential states to algebraic ones by setting associated parameters to zero.
\section{Mass-Matrix Explicit DAE Formulation}
\subsection{Mass-Matrix DAE Formulation}
The proposed mass-matrix formulation is given by
\begin{equation}
    \label{eq:mass-matrix-dae}
    \begin{array}{rcl}
    \underbrace{
    \begin{bmatrix}
    \bm{M_x} & \bm{0} \\
    \bm{0} & \bm{0}
    \end{bmatrix}
    }_{\bm{M}}
    \begin{bmatrix}
    \dot{\bm{x}}
    \\ 
    \bm{0}
    \end{bmatrix}
    &=&
    \begin{bmatrix}
    \bm{\hat{f}}(\bm{x}, \bm{y}, \bm{u}) \\
    \bm{g}(\bm{x}, \bm{y}, \bm{u})
    \end{bmatrix}
    \end{array} \, , 
\end{equation}
where $\bm{M}$ is a $(n+m) \times (n+m)$ possibly-diagonal matrix with its upper-left $n \times n$ block being $\bm{M_x}$.
$\bm{\hat{f}}$
($\bm{\hat{f}}: \mathbb{R}^{m+n+o} \Rightarrow \mathbb{R}^n$)
is chosen based on $\bm{f}$ and determines the diagonality of $\bm{m}$.

The choice of $\bm{\hat{f}}$ is not unique. 
For transient stability simulations, the simplest choice of $\bm{\hat{f}}$ is the numerators of 
$\bm{f}$ if $\bm{f}$ is a fraction expression with a constant parameter as the denominator. In such case, $\bm{M_x}$ becomes a time-invariant diagonal matrix with the element $\mu_{i, i}$ $(i \in [1, n])$ being 
\begin{equation}
\begin{array}{rcl}
\mu_{i, i} &=& \hat{f}_i / f_i %
\end{array}
\, .
\end{equation}

The proposed mass-matrix formulation in \eqref{eq:mass-matrix-dae} has the following relationships with the existing ones:
\begin{enumerate}
    \item \eqref{eq:mass-matrix-dae} is a special case of \eqref{eq:dae-semi-implicit} when one let $\bm{\tilde{f}} = \bm{\hat{f}}$, $\bm{T(x, y)} = \bm{M_x}$ and $\bm{R(x, y)} = \bm{0}$.
    \item \eqref{eq:dae-flexible} is a special case of \eqref{eq:mass-matrix-dae} when one let $\bm{\hat{f}} = \bm{f}$ and $\bm{M} = \bm{\Gamma}$.
    \item \eqref{eq:dae-traditional} is a special case of \eqref{eq:dae-flexible} when one let $\bm{\Gamma} = \bm{I_n}$, an $n \times n$ identity matrix.
    Due to the fact that the traditional formulation has already been proven to be all-encompassing for power system models, \eqref{eq:mass-matrix-dae}, as a superset of \eqref{eq:dae-traditional}, can naturally accommodate for all power system equation formulations.

\end{enumerate}
Such relationships can be represented as:
\begin{equation}
    \text{\{Semi-Implicit\}} \supset \text{\{\textbf{Mass-Matrix}\}} \supset \text{\{Flexible\}} \supset \text{\{Traditional\}}
\end{equation}

The main advantages of \eqref{eq:mass-matrix-dae} are two-fold:
\begin{enumerate}
    \item As a general formulation, it is directly supported by a broad spectrum of full-fledged DAE solvers. Power system researchers can thus focus on formulating models without worrying about the details in solution techniques.
    \item Simple to implement. The only required change to the traditional formulation is to separate the parameters (typically, time or mass constants) from the denominators of $\bm{f}$ into the mass matrix. 
    Mixing models in the mass-matrix and the traditional formulations is allowed, as the latter is a special case.
\end{enumerate}

In addition, the mass-matrix formulation shares the same advantage as the semi-implicit formulation that allows model simplification by setting some time constants to zero.

\subsection{Modeling Examples}
This section presents some basic transfer functions and synchronous generators in the mass-matrix formulation, in order to demonstrate the implementation simplicity.

\subsubsection{First-order lag block}
The conventional formulation of lag block has one differential equations given by
\begin{equation}
\begin{array}{rll}
    \dot{y} &=& (Ku - y) / T \, .
\end{array}
\end{equation}
Using the proposed formulation, the numerator can be chosen as the right-hand-side (RHS) by moving  $T$ to the mass matrix diagonal:
\begin{equation}
\begin{array}{rll}
    T \dot{y} &=& (Ku - y) \, ,
\end{array}
\end{equation}
which allows using $T=0$ to convert the lag block to a pure gain.
This formulation is the same as using the semi-implicit formulation.
\subsubsection{Lead-lag block}
The lead-lag block can be implemented in the following serial approach:
\begin{equation}
\begin{array}{rllll}
    \dot{x'} &=& (u - x') / T_2 \\
    y &=& (T_1/T_2)  (u - x') + x'  \, .
\end{array}
\end{equation}
The mass-matrix formulation can be given by
\begin{equation}
\begin{array}{rll}
    T_2 \dot{x'} &=& (u - x') \\
    0 &=& T_1 T_2'  (u - x') + x' - y  \, ,
\end{array}
\end{equation}
where $T_2'$ is an auxiliary parameter given by
\begin{equation}
\left\{\begin{array}{llll}
T_2' &=& 1/T_2, & \text{if } T_2 \ne 0 \\ 
T_2' &=& 0, & \text{otherwise} \, .
\end{array}\right.
\end{equation}
$T_2'$ can be pre-calculated before simulation and thus will not increase the computation operations. This implementation retains the ability to convert the lead-lag (when $T_1 = 0$) to a lag block or to a pass-through block (when $T_1 = T_2 = 0$).
\subsubsection{Synchronous Generator Model}
The round-rotor generator model \cite{chow2020power} is conventionally given by 
\begin{equation}
\begin{array}{rll}
    \dot{e}'_{q} &=& (- X_{ad}I_{fd} + v_{f}) / {T'_{d0}} \\ 
    \dot{e}'_{d} &=& (- X_{aq}I_{1q}) / {T'_{q0}} \\
    \dot{e}''_{d} &=& (- I_{d} \left(x'_{d} - x_{l}\right) - e''_{d} + e'_{q}) / {T''_{d0}} \\ 
    \dot{e}''_{q} &=& (I_{q} \left(x'_{q} - x_{l}\right) - e''_{q} + e'_{d}) / {T''_{q0}}
\end{array}
\end{equation}
where $X_{ad}I_{fd}$ and $X_{aq}I_{1q}$ are calculated algebraically. 
The proposed mass-matrix DAE formulation is 
\begin{equation}
\begin{array}{rll}
    T'_{d0} \dot{e}'_{q} &=& - X_{ad}I_{fd} + v_{f}  \\ 
    T'_{q0} \dot{e}'_{d} &=& - X_{aq}I_{1q}   \\
    T''_{d0} \dot{e}''_{d} &=& - I_{d} \left(x'_{d} - x_{l}\right) - e''_{d} + e'_{q}  \\ 
    T''_{q0} \dot{e}''_{q} &=& I_{q} \left(x'_{q} - x_{l}\right) - e''_{q} + e'_{d}
\end{array}
\end{equation}

Generator model reductions can be achieved by setting the corresponding time constants to zero. For example, the one d- and one q-axis flux-decay model can be obtained by setting $T_{d0}'' = T_{q0}'' = 0$.

As seen from the examples, \eqref{eq:mass-matrix-dae} is straightforward to implement.
In a symbolic modeling framework where parameters, variables, and equations are symbols \cite{cui2020hybrid}, the conversion from the traditional formulation can be automated by manipulating the equations with a single parameter as the denominator.

\subsection{Implicit Trapezoidal Method for Mass-Matrix DAE}
The mass-matrix formulation of DAE can be solved by a variety of numerical integration methods.
Implicit numerical integration schemes are known for their good performance for Ordinary Differential Equations (ODEs) and DAEs that are stiff, which is the case for transient simulation.
At each step, ITM solves a set of nonlinear equations consisting of both differential and algebraic equations simultaneously through Newton's iteration. 
This subsection discusses the form of iteration for the mass-matrix formulation.

The nonlinear equations to solve for \eqref{eq:dae-traditional} at time $t$ are given by 
\begin{equation}
\begin{array}{rrcll}
     \bm{0} &=& \bm{p}_t &=& (\bm{x}_t - \bm{x}_{t-h}) - 0.5 h (\bm{f}_t + \bm{f}_{t-h} ) \\
     \bm{0} &=& \bm{q}_t &=& -\gamma \bm{g}_t 
\end{array} \, ,
\end{equation}
where $h$ is the step size, and $\bm{x}_{t-h}$  and $\bm{f}_{t-h}$ are the differential states and equation RHS computed at the previous time ($t-h$).
Note that for $\bm{g}$, the sign and the scaling factor $\gamma$ can be chosen arbitrarily, and a small $\gamma$ close to $h$ improves the convergence.
Accordingly, solutions are obtained by iteratively updating variables using \eqref{eq:itm-variable}, where the increments are calculated using \eqref{eq:itm-newton-inner}.
\begin{equation}
    \label{eq:itm-variable}
    \begin{array}{rcl}
    \begin{bmatrix}
    \bm{x}^{(i+1)} (t)
    \\ 
    \bm{y}^{(i+1)} (t)
    \end{bmatrix}
    &=&
    \begin{bmatrix}
    \bm{x}^{(i)} (t)
    \\ 
    \bm{y}^{(i)} (t)
    \end{bmatrix}
    + 
    \begin{bmatrix}
    \Delta\bm{x}^{(i)} 
    \\ 
    \Delta \bm{y}^{(i)}
    \end{bmatrix}
    \end{array}
\end{equation}
\begin{equation}
    \label{eq:itm-newton-inner}
    \begin{array}{rcl}
    \begin{bmatrix}
    \Delta\bm{x}^{(i)} 
    \\ 
    \Delta \bm{y}^{(i)}
    \end{bmatrix}
    &=&
    -
    {\underbrace{
    \begin{bmatrix}
    \bm{I}_n - 0.5h \bm{f_x}^{(i)} & -0.5 h \bm{f_y}^{(i)} \\
    -\gamma \bm{g_x}^{(i)} & -\gamma \bm{g_y}^{(i)}
    \end{bmatrix}
    }_{\bm{A}^{(i)}}}^{-1}
    \begin{bmatrix}
    \bm{p}^{(i)} \\
    \bm{q}^{(i)}
    \end{bmatrix}
    \end{array}
\end{equation}

Comparing \eqref{eq:mass-matrix-dae} with \eqref{eq:dae-traditional}, we can observe that the mass-matrix formulation can be derived by multiplying $\bm{M}$ to both sides of the equation.
Therefore, mutltiplying $\bm{M}$ to \eqref{eq:itm-newton-inner} yields
\begin{equation}
\begin{array}{rrcll}
     \bm{0} &=& \bm{\hat{p}}_t &=& \bm{M_{x}} (\bm{x}_t - \bm{x}_{t-h}) - 0.5 h (\bm{\hat{f}}_t + \bm{\hat{f}}_{t-h} ) \\
     \bm{0} &=& \bm{q}_t &=& - \gamma \bm{g}_t 
\end{array} \, .
\end{equation}
The Jacobian matrix for calculating increments is given by:
\begin{equation}
    \label{eq:mass-matrix-jacobian}
    \begin{array}{rcl}
    \bm{\hat{A}}^{(i)}
    &=&
    \begin{bmatrix}
    \bm{M_x} - 0.5h \bm{\hat{f}_x}^{(i)} & -0.5 h \bm{\hat{f}_y}^{(i)} \\
    - \gamma \bm{g_x}^{(i)} & - \gamma \bm{g_y}^{(i)}
    \end{bmatrix}
    \end{array} \, .
\end{equation}
Comparing \eqref{eq:mass-matrix-jacobian} to \eqref{eq:itm-newton-inner}, one can notice that the only required change to the solver is to substitute the identity matrix $\bm{I_n}$ with the mass matrix $\bm{M_x}$. 
Therefore, numerical integration routines can be readily adapted for the proposed formulation.

\subsection{Remarks on the Computational Complexity}
In terms of computational complexity, \eqref{eq:mass-matrix-dae} is not more demanding than \eqref{eq:dae-traditional}.
When one uses a constant diagonal mass-matrix formulation, the number of division operations can be slightly reduced for evaluating $\bm{\hat{f}}$ and the subsequent Jacobian elements.
Such effect, however, is negligible in actual test cases mostly because the number of reduced operations is small w.r.t. the total operations.

\section{Case Studies}
\subsection{Solvers and Simulation Setup}

The most significant advantage of the mass-matrix formulation is its compatibility with fine-tuned DAE solver packages, which usually come with a multitude of solution methods and error control mechanisms and have undergone rigorous tests.
Such compatibility allows power system researchers to focus on modeling while utilizing the state-of-the-art numerical solvers.

\begin{table}
\centering
\caption{An overview of the tested solvers}
\label{tab:solver-table}
\begin{tabular}{llccl} 
\toprule
 & \textit{\textbf{Solver Name}}  & \textit{\textbf{Order}}  & \textit{\textbf{Stability}}  & \textit{\textbf{Remarks}}                                                                         \\ 
\midrule
\multicolumn{5}{l}{\textit{Rosenbrock methods (for small-stiffness problems)} }                                                                                                                 \\
 & Ros4LStab                      & 4                        & A                            &                                                                                                   \\ 
\cdashline{2-5}[1pt/1pt]
 & Rodas4                         & 4                        & A                            & Order 3 interpolant                                                                               \\ 
\cdashline{2-5}[1pt/1pt]
 & Rodas42                        & 4                        & A                            & Order 3 interpolant                                                                               \\ 
\cdashline{2-5}[1pt/1pt]
 & Rodas4P                        & 4                        & A                            & \begin{tabular}[c]{@{}l@{}}Order 3 interpolant~with\\parabolic~correction \end{tabular}           \\ 
\cdashline{2-5}[1pt/1pt]
 & Rodas5                         & 5                        & A                            & Hermite interpolant                                                                               \\ 
\midrule
\multicolumn{5}{l}{\textit{Rosenbrock-Wanner methods (allow approximate Jacobian)} }                                                                                                            \\
 & Rosenbrock23                   & 2/3                      & L                            &                                                                                                   \\ 
\cdashline{2-5}[1pt/1pt]
 & ROS34PW1a                      & 4                        & L                            &                                                                                                   \\ 
\cdashline{2-5}[1pt/1pt]
 & ROS34PW1b                      & 4                        & L                            &                                                                                                   \\ 
\cdashline{2-5}[1pt/1pt]
 & ROS34PW2                       & 4                        & L                            &                                                                                                   \\ 
\cdashline{2-5}[1pt/1pt]
 & ROS34PW3                       & 4                        & A                            &                                                                                                   \\ 
\midrule
\multicolumn{5}{l}{\textit{Implicit Runge-Kutta methods (for stiff problems, low accuracy)} }                                                                                                   \\
 & Trapezoid                      & 2                        & A                            & Adaptive time step                                                                                \\ 
\cdashline{2-5}[1pt/1pt]
 & ImplicitEuler                  & 1                        & L                            & Adaptive time step                                                                                \\ 
\hline
\multicolumn{5}{l}{\textit{Multistep methods (for stiff problems)} }                                                                                                                            \\
 & QNDF1                          & 1                        & L                            & Quasi-constant time step                                                                          \\ 
\cdashline{2-5}[1pt/1pt]
 & QNDF2                          & 2                        & L                            & Quasi-constant time step                                                                          \\ 
\cdashline{2-5}[1pt/1pt]
 & QBDF1                          & 1                        & L                            & \begin{tabular}[c]{@{}l@{}}Equivalent to ImplicitEuler \\with BDF error estimator \end{tabular}  \\ 
\cdashline{2-5}[1pt/1pt]
 & QBDF2                          & 2                        & L                            &                                                                                                   \\ 
\cdashline{2-5}[1pt/1pt]
 & QNDF                           & 1$\sim$5                 & L                            &                                                                                                   \\
\bottomrule
\end{tabular}
\end{table}

For example, the \lstinline{DifferentialEquations.jl} package \cite{rackauckas2017differentialequations} provides four categories of methods that naturally support the mass-matrix formulation: Rosenbrock methods and Rosenbrock-Wanner methods (for small stiffness systems), and Implicit Runge-Kutta methods and multistep methods (for stiff problems).
An overview of the interfaced solvers in this work is shown in \Cref{tab:solver-table}, where ``A" or ``L" in the stability column indicates an algorithm being A-stable or L-stable \cite{wanner1996solving}.
Details of the solvers can be found in \cite{rackauckas2017differentialequations} and the accompanying documentation.

The proposed work has been implemented in ANDES \cite{cui2020hybrid}, a Python-based hybrid symbolic-numeric library for power systems, and interfaced with the solvers in the Julia language with the Jacobian callback provided.
Case studies utilize Python 3.7.6, ANDES 1.0.8, Julia 1.5.0, and DifferentialEquations 6.15.0. 
Simulations are executed on an Intel Xeon W-2133 CPU with 32 GiB of RAM running Ubuntu 16.04 LTS.

\subsection{Solver Benchmarks and Statistics}

\begin{figure}[!t]
\centering
\includegraphics[width=\columnwidth]{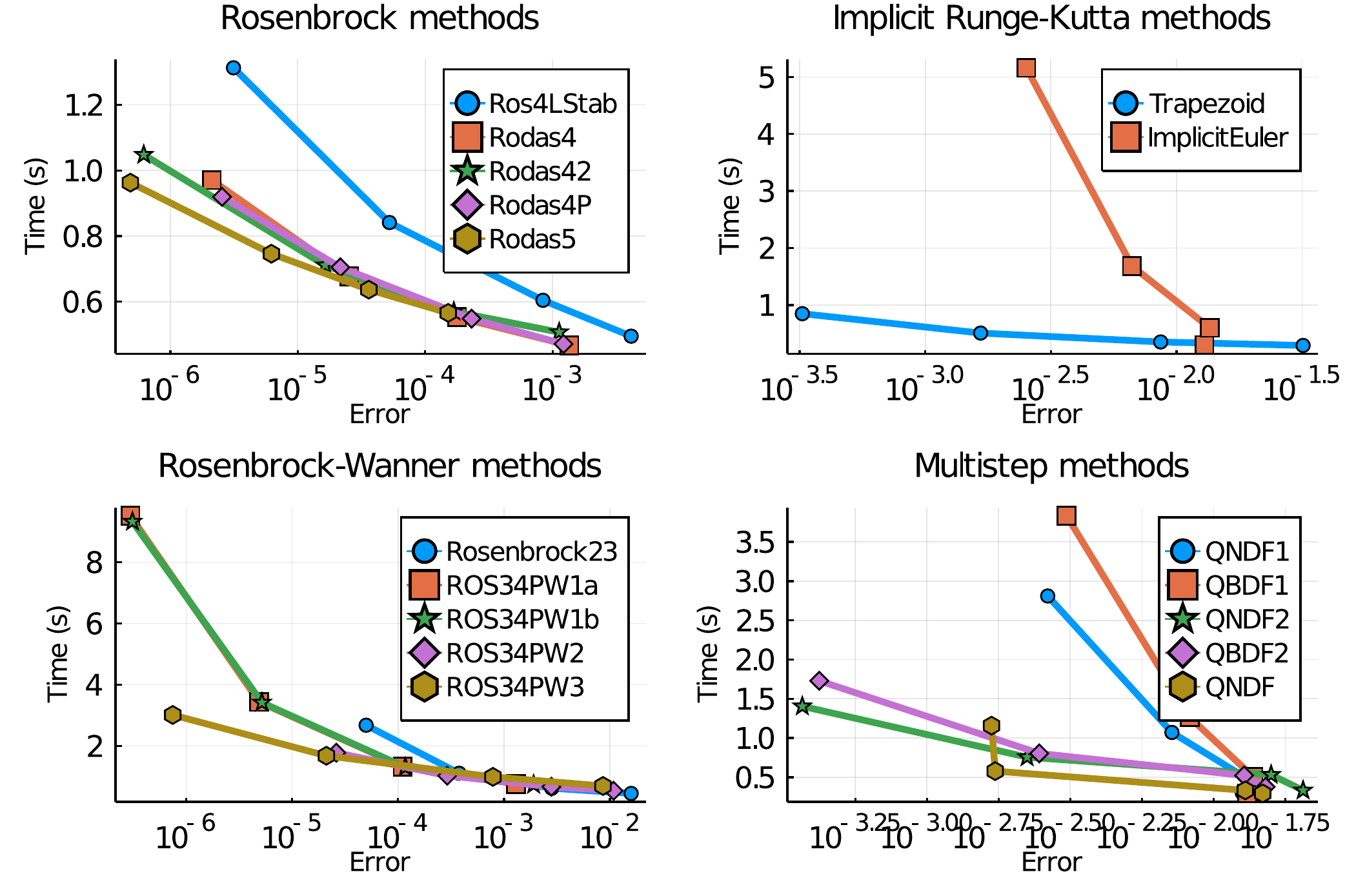}
\caption{Performance of the solvers using the Kundur's system.}
\label{fig:solver-error-kundur}
\end{figure}

First, the Kundur's system modeled in the proposed formulation is used for solver benchmarking. 
The system contains 52 differential states and 140 algebraic variables, and the rank of the mass matrix is 48 (two generators are reduced with $T_{d0}'' = T_{q0}'' = 0$).
The response following a line trip at $t=0.1$ s and a reconnection after 50 ms is simulated for 5 s.
The baseline ``accurate" solution is obtained using \lstinline{Rodas5}, which is efficient in high accuracy, with both absolute and relative tolerances set to $10^{-12}$.
Next, each solver is given four pairs of absolute and relative tolerances, chosen from $(10^{-5}, 10^{-6}, 10^{-7}, 10^{-8})$, and $(10^{-1}, 10^{-2}, 10^{-3}, 10^{-4})$, respectively.
Errors are obtained as the difference between the accurate and the actual solutions at the final simulation step.
Each case is run for five times to compute the average time. 
The results shown in \figref{fig:solver-error-kundur}, and some observations are:
\begin{enumerate}
    \item The commonly used Trapezoidal method is balanced in speed and accuracy. ImplicitEuler is neither efficient nor accurate.
    \item Depending on the stiffness, the Rosenbrock methods can be faster than the Trapezoid method at some accuracy levels.
    \item The Rosenbrock-Wanner methods allow approximate Jacobians. When accurate Jacobians are used (in our case and most other transient stability simulation tools), the Rosenbrock-Wanner methods tend to be slower than the Rosenbrock ones.
    \item The QNDF method has similar speed and accuracy to the Trapezoid method. QBDF1 performs similarly to Implicit Euler.
\end{enumerate}

\begin{figure}[!t]
\centering
\includegraphics[width=\columnwidth]{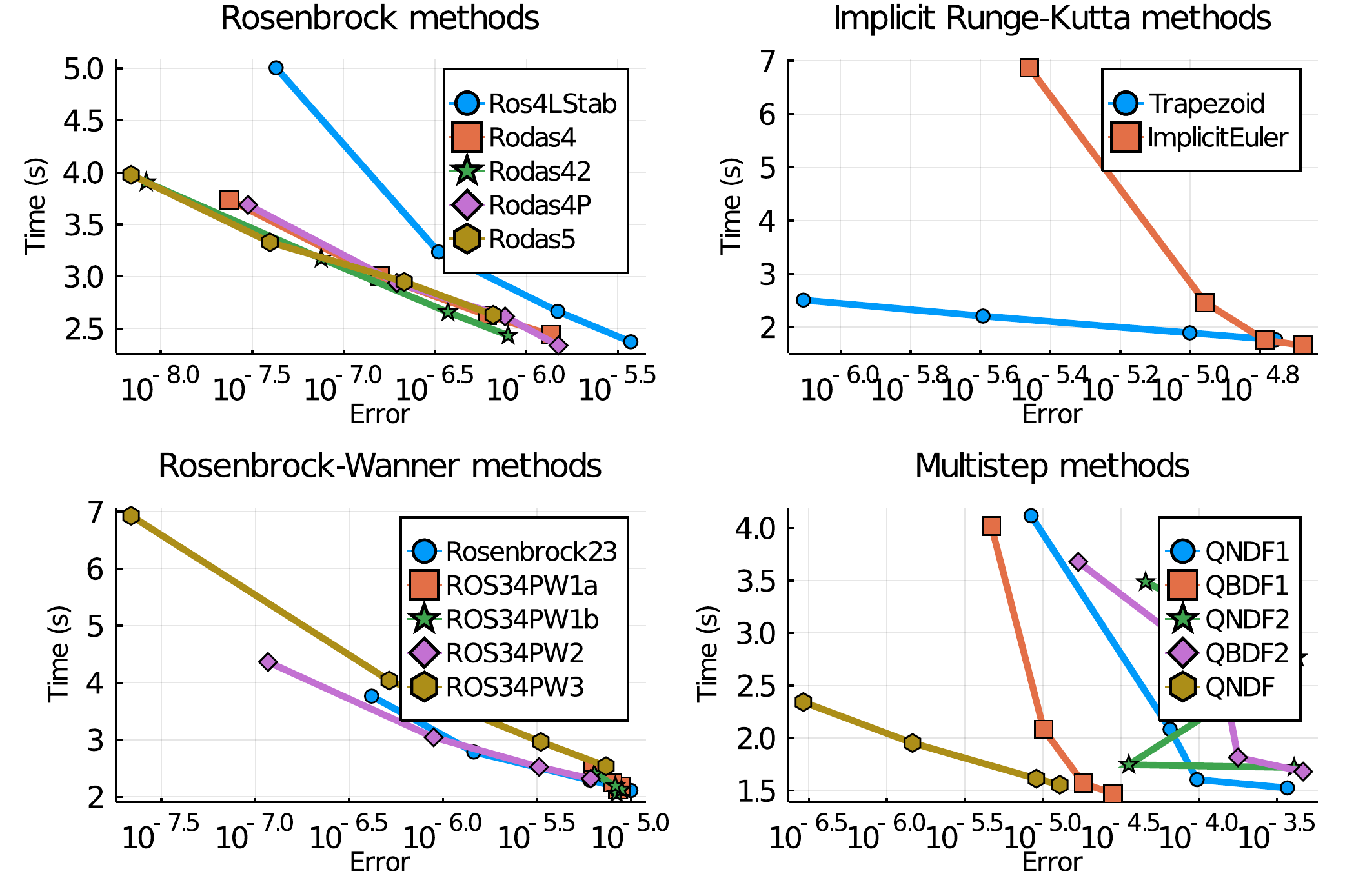}
\caption{Performance of the solvers using the GB 2,224-bus system.}
\label{fig:solver-error-2000}
\end{figure}

Next, the Great Britain (GB) 2,224-bus system \cite{gbnetwork2020} with dynamics is used for benchmarking.
The system contains 788 state variables and 9,176 algebraic variables. 
A disturbance of line trip at $t=1.0$ s and a reconnection after 100 ms is simulated for 5 s.
Benchmark results are shown in \figref{fig:solver-error-2000}. 
The observations from the Kundur's system also apply to this case.
For a different system, one can perform similar benchmarks to identify the best solvers that satisfy the speed and accuracy requirements.  

\section{Conclusions}
\label{sec:conclusions}

This letter proposes a mass-matrix formulation for transient simulation with the advantages of being compatible 
with the traditional formulation, compatible with a multitude of solvers, and simple to implement.
Modeling examples for common transfer functions and synchronous generators are shown,
and the implicit trapezoid method is deduced for solving DAE in the mass-matrix formulation.
The solver compatibility is verified using 17 solvers benchmarked for two test systems.
In conclusion, the compatibility and simplicity make the proposed method highly suitable for transient stability simulation.

\ifCLASSOPTIONcaptionsoff
  \newpage
\fi

\bibliographystyle{IEEEtran}
\bibliography{IEEEabrv,papers}

\begin{thebibliography}{1}
\providecommand{\url}[1]{#1}
\csname url@samestyle\endcsname
\providecommand{\newblock}{\relax}
\providecommand{\bibinfo}[2]{#2}
\providecommand{\BIBentrySTDinterwordspacing}{\spaceskip=0pt\relax}
\providecommand{\BIBentryALTinterwordstretchfactor}{4}
\providecommand{\BIBentryALTinterwordspacing}{\spaceskip=\fontdimen2\font plus
\BIBentryALTinterwordstretchfactor\fontdimen3\font minus
  \fontdimen4\font\relax}
\providecommand{\BIBforeignlanguage}[2]{{%
\expandafter\ifx\csname l@#1\endcsname\relax
\typeout{** WARNING: IEEEtran.bst: No hyphenation pattern has been}%
\typeout{** loaded for the language `#1'. Using the pattern for}%
\typeout{** the default language instead.}%
\else
\language=\csname l@#1\endcsname
\fi
#2}}
\providecommand{\BIBdecl}{\relax}
\BIBdecl

\bibitem{aristidou2013dynamic}
P.~Aristidou, D.~Fabozzi, and T.~Van~Cutsem, ``Dynamic simulation of
  large-scale power systems using a parallel schur-complement-based
  decomposition method,'' \emph{IEEE Transactions on Parallel and Distributed
  Systems}, vol.~25, no.~10, pp. 2561--2570, 2013.

\bibitem{milano2016semi}
F.~Milano, ``Semi-implicit formulation of differential-algebraic equations for
  transient stability analysis,'' \emph{IEEE Transactions on Power Systems},
  vol.~31, no.~6, pp. 4534--4543, 2016.

\bibitem{chow2020power}
J.~H. Chow and J.~J. Sanchez-Gasca, \emph{Power System Modeling, Computation,
  and Control}.\hskip 1em plus 0.5em minus 0.4em\relax John Wiley \& Sons,
  2020.

\bibitem{cui2020hybrid}
H.~Cui, F.~Li, and K.~Tomsovic, ``Hybrid symbolic-numeric library for power
  system modeling and analysis,'' \emph{arXiv preprint arXiv:2002.09455}, 2020.

\bibitem{rackauckas2017differentialequations}
C.~Rackauckas and Q.~Nie, ``Differentialequations. jl--a performant and
  feature-rich ecosystem for solving differential equations in julia,''
  \emph{Journal of Open Research Software}, vol.~5, no.~1, 2017.

\bibitem{wanner1996solving}
G.~Wanner and E.~Hairer, \emph{Solving ordinary differential equations
  II}.\hskip 1em plus 0.5em minus 0.4em\relax Springer Berlin Heidelberg, 1996.

\bibitem{gbnetwork2020}
\BIBentryALTinterwordspacing
T.~U. of~Edinburgh. (2020, aug) Gb network. [Online]. Available:
  \url{https://www.maths.ed.ac.uk/optenergy/NetworkData/fullGB/}
\BIBentrySTDinterwordspacing

\end{thebibliography}

\end{document}